\documentclass{phb-proc4-auth}

\usepackage{graphicx}
\usepackage{amssymb,amsmath}

\begin{document}
\begin{frontmatter}


\journal{SCES '04}


\title{ Mean--field phase diagram of interacting $e_{g}$ electrons }


\author [FR,PL]{Marcin Raczkowski,\corauthref{cor}\thanksref{KBN}}
\corauth[cor]{Corresponding author (Marcin Raczkowski).\newline
{\it Email address:} m.raczkowski@if.uj.edu.pl}
\author[PL]{Andrzej M. Ole\'s}
\author[FR]{Raymond Fr\'esard}

%
 
\address[FR]{Laboratoire CRISMAT, UMR CNRS--ENSICAEN(ISMRA) 6508,
6 Bld. du Mar\'echal Juin, F-14050 Caen, France}
\address[PL]{Marian Smoluchowski Institute of Physics, Jagellonian
 University, Reymonta 4, PL-30059 Krak\'ow, Poland}

%

\thanks[KBN]{M.R. was supported by a Marie Curie fellowship of the
             European Community program under number HPMT2000--141.
             This work was supported by the Polish State Committee of
             Scientific Research (KBN) under Project No. 1~P03B~068~26.}

%
%
%
%


\begin{abstract}
We investigate the magnetic phase diagram of the two-dimensional model 
for $e_g$ electrons which describes layered nickelates. One finds a 
generic tendency towards magnetic order accompanied by orbital 
polarization. For two equivalent orbitals with diagonal hopping such 
orbitally polarized phases are induced by finite crystal field.

\end{abstract}

%
%



\end{frontmatter}

%

Correlated $e_g$ electrons lead to several interesting phenomena 
in cuprates and manganites. At large intraorbital Coulomb interaction 
$U$ and finite Hund's exchange $J_H$, the phase diagram at the filling 
$n=1$ electron per site shows an interesting competition between 
ferromagnetic (FM) and antiferromagnetic (AF) instabilities \cite{Fei97}. 
Recent studies have further emphasized the role of $J_H$ at orbital 
degeneracy \cite{ddHH}; here we address this problem starting from 
weak coupling. 

We consider a two-dimensional (2D) model, 
\begin{equation}
H=H_0+H_{\mathnormal int}, 
\end{equation}
for $e_g$ electrons, with the one-particle term:
\begin{equation}
 H_0 =     \sum_{\langle ij\rangle}\sum_{\alpha\beta\sigma} 
           t^{\alpha\beta}_{ij}
           c^{\dag}_{i\alpha\sigma}c^{}_{j\beta\sigma} 
         + \tfrac{1}{2}E_z\sum_{i} (n_{ix}-  n_{iz}).
\end{equation}
%
For $e_g$ orbitals: $|x\rangle\sim|x^{2}-y^{2}\rangle$ and 
$|z\rangle\sim|3z^{2}-r^{2}\rangle$, one has the hopping matrix
$t^{\alpha\beta}_{ij}=-\tfrac{t}{4}\bigl( 
\begin{smallmatrix}
3           & \pm\sqrt{3} \\
\pm\sqrt{3} &  1
\end{smallmatrix}\bigr)$, 
where $t$ stands for a $(dd\sigma)$ hopping matrix element, and the 
off--diagonal hopping along $a$ and $b$ axis depends on the phases 
of the $|x\rangle$ orbital. The crystal field splitting $\propto E_z$, 
{\it a priori} finite in a 2D model, lifts the degeneracy of $e_g$ 
orbitals. We compare this case with the frequently studied 
{\it diagonal-hopping model} \cite{ddHH}, 
$t^{\alpha\beta}_{ij}=-\tfrac{t}{2}\delta_{\alpha\beta}$, 
i.e., with the degenerate Hubbard model. 

Coulomb interactions $H_{\mathnormal int}$ are described by
on-site intraorbital Coulomb $U$ and exchange $J_H$ elements \cite{Amo83}; 
here we write the leading terms using the operators: 
$n_i = \sum_{\alpha\sigma}n_{i\alpha\sigma}$, 
$m_i = \sum_{\alpha\sigma}\lambda_{\sigma} n_{i\alpha\sigma}$, 
$o_i = \sum_{\alpha\sigma}\lambda_{\alpha} n_{i\alpha\sigma}$, and 
$p_i = \sum_{\alpha\sigma}\lambda_{\alpha}
                          \lambda_{\sigma} n_{i\alpha\sigma}$, 
with $\lambda_{\alpha}=\pm 1$ for $\alpha=x(z)$ orbital, and 
     $\lambda_{\sigma}=\pm 1$ for $\sigma=\uparrow(\downarrow)$ spin: 
\begin{eqnarray}
H_{\mathnormal int} &=& \tfrac{1}{8}\sum_{i}\big[ 
               (3U-5J_H) n_{i}^2 - (U+J_H) m_{i}^2 \nonumber \\
& &\hskip .8cm -(U-5J_H) o_{i}^2 - (U-J_H) p_{i}^2 \big].
\end{eqnarray}
For the electron density $n<2$ (doping $x=2-n$), the model (1) 
describes layered La$_{2-x}$Sr$_{x}$NiO$_4$ nickelates. 
The coefficient $(U+J_H)$ of the $m_i^2$ term is the Stoner parameter. 

\begin{figure}[t!]
\begin{center}
\includegraphics[scale=0.44]{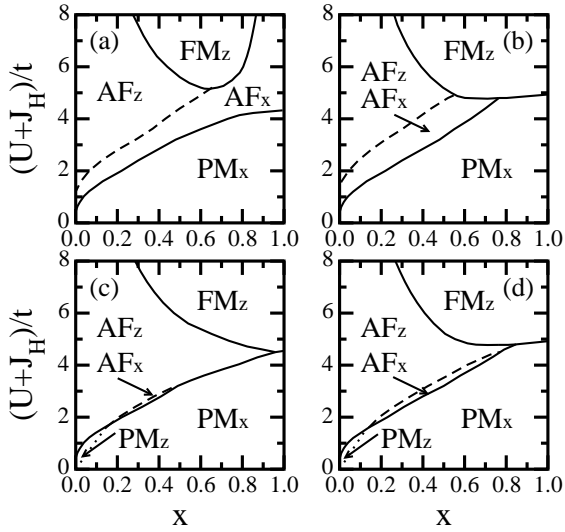}
\end{center}
\caption
{
Phase diagrams of the $e_g$ orbital model (1) as functions of the 
Stoner parameter $U+J_H$ and hole doping $x=2-n$, with $E_z=0$ (a,b) 
and $E_z=0.1U$ (c,d). Left (right) panels show results for
$J_H=0.15U$ ($J_H=0.25U$). Dashed (dotted) line indicates the 
AF$x$--AF$z$ (PM$x$--PM$z$) phase transition. 
}
\label{fig1}
\end{figure}

We investigate the stability of various phases with either uniform or
staggered magnetic order by tackling the above Hamiltonian in mean-field
approximation on a 128$\times$128 cluster, using periodic boundary 
conditions at low temperature $T=0.01t$, for two 
representative values of Hund's exchange: $J_H=0.15U$ and $J_H=0.25U$. 
One finds that the paramagnetic (PM) phase can be unstable towards: 
($i$) orbitally polarized paramagnetic (PM${\alpha}$) phase 
(with the enhanced electron density in $\alpha$ orbitals); 
($ii$) FM or orbitally polarized FM (FM${\alpha}$) phases; 
($iii$) AF or orbitally polarized AF (AF${\alpha}$) ones. 
The phases with orbital polarization are generic for the 2D model of
$e_g$ electrons (Fig.~\ref{fig1}) -- they are characterized by finite 
values of the $\langle o\rangle$ and possibly $\langle p\rangle$ order 
parameters. For instance, both types of AF phases, AF$x$ with 
$\langle o\rangle>0$ (the system favors $|x\rangle$ over $|z\rangle$ 
occupancy), and AF$z$, compete with each other and are stable in 
different regions.   

In the intermediate correlated regime ${U+J_H}\simeq 2 t$ the AF$x$ 
phase is stabilized due to the larger hopping element between 
$|x\rangle$ orbitals, whereas in the strongly correlated regime 
${U+J_H}\geqslant 5t$, the system favors instead $|z\rangle$ orbitals 
which allows to better optimize the magnetic energy owing to a negative 
$\langle p\rangle$. However, upon increasing $x$ the double occupancy 
energy becomes less important compared to a possible kinetic energy 
gain when $|x\rangle$ orbitals are closer to half filled. This leads 
to a first order phase transition at $x\gtrsim 0.6$, with all the 
order parameters changing discontinuously. 

In a similar way one can understand the origin of PM$x$ state 
(optimizing the kinetic energy) and the stability of FM$z$ phase 
in the strongly correlated regime at finite doping (optimizing the 
magnetic energy). The latter phase is strongly favored by a large value 
of $J_H=0.25U$, so that the usual AF instability at $x=1$ is fully 
suppressed. Moreover, at $J_H>0.2U$ the attractive interaction in the 
$o$ channel changes into a repulsive one. As a consequence, the energy 
of the AF$x$ phase cannot be much reduced by a large positive 
$\langle o\rangle$, and electrons redistribute almost equally over 
the two orbitals. Therefore, double occupancies might be avoided 
{\it even without} creating a finite magnetic moment which makes the 
PM$x$ region broader for $J_H=0.25U$ than for $J_H=0.15U$. We note 
that the realistic value of $E_z=0.1U$, as reported for La$_2$NiO$_4$ 
\cite{Kuip}, has only little influence on the phase diagram, except for 
strong reduction of the region of stability of the AF$x$ phase. 
Remarkably, the FM$z$--PM$x$ line [Fig.~\ref{fig1}(d)] is nearly 
unaltered by finite $E_z$, showing that the FM instability almost 
decouples from the orbital polarization (when both $\langle o\rangle$ 
and $\langle p\rangle$ order parameters are suppressed). 

\begin{figure}[t!]
\begin{center}
\includegraphics[scale=0.44]{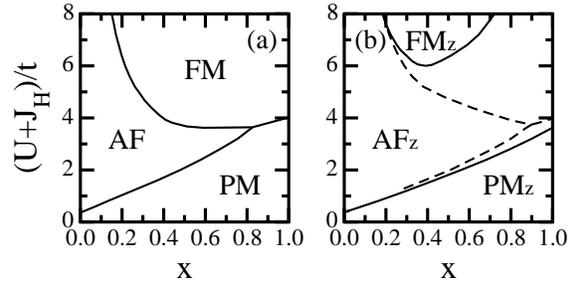}
\end{center}
\caption
{
Phase diagrams as in Fig. 1 for the degenerate Hubbard model, for: 
(a) $E_z=0$ and $J_H=0.25U$; 
(b) $E_z=0.1U$, with $J_H=0.15U$ (solid line) and $J_H=0.25U$ (dashed line).
}
\label{fig2}
\end{figure}

Finally, we investigate the magnetic instabilities of the degenerate 
Hubbard model in the subspace with $\langle o\rangle=\langle p\rangle=0$ 
[Fig.~\ref{fig2}(a)]. In qualitative agreement with Ref. \cite{Amo83}, 
the instabilities of the PM phase towards an AF/FM one for small/large 
doping depend only on the Stoner parameter $U+J_H$; they were also 
found in the strongly correlated regime \cite{Web98}. The phase 
diagram at finite $E_z$ [Fig.~\ref{fig2}(b)] depends on $J_H$ in a 
non-trivial way, and we recover the AF$z$ phase in the vicinity of
$x=1$, provided $J_H$ is small enough. 
Namely, when one orbital is sufficiently favored over the other one, 
the $S=\tfrac{1}{2}$ AF$z$ phase is obtained, since some processes 
of the FM superexchange are blocked. 
  
Summarizing, we have found that both the realistic $e_g$ hopping 
$t^{\alpha\beta}_{ij}$ and finite crystal field splitting $E_z$ act to 
stabilize {\it orbitally polarized phases}. The phase diagram of the $e_g$ 
electron model is richer than that of the degenerate Hubbard model 
\cite{Amo83}, and favors AF phases. For instance, at a fixed value of 
the Stoner parameter $U+J_H$, the crossover from the AF to FM phase 
occurs for $e_g$ electrons at higher doping $x$.

%



\begin{thebibliography}{00}

\bibitem{Fei97} L.F. Feiner, A.M. Ole\'s, and J. Zaanen, 
                  Phys. Rev. Lett. 78 (1997) 2799.

\bibitem{ddHH} A. Klejnberg and J. Spa\l{}ek, 
                  Phys. Rev. B 61 (2000) 15542;
               S. Florens {\it et al.}, 
                  {\it ibid.} 66 (2002) 205102; 
	       Y. \={O}no, M. Potthoff, and R. Bulla, 
                  {\it ibid.} 67 (2003) 035119;
	       A. Koga {\it et al.}, 
	          Phys. Rev. Lett. 92 (2004) 216402. 	  
 
\bibitem{Amo83} A.M. Ole\'s, 
                  Phys. Rev. B 28 (1983) 327.

\bibitem{Kuip} P. Kuiper {\it et al.}, Phys. Rev. B 44 (1991) 4570;
               P. Kuiper {\it et al.}, {\it ibid.} 57 (1998) 1552.

\bibitem{Web98} J. B\"unemann, W. Weber, and F. Gebhard,
                  Phys. Rev. B 57 (1998) 6896.

\end{thebibliography}
\end{document}